# A method for fabricating a micro-structured surface of polyimide with open and closed pores

**Yong-Won Ma**[a], Jae Yong Oh[b], Seokyoung Ahn[c], Bo Sung Shin[a,d,*]

[a]*Department of Cogno-Mechatronics Engineering, Pusan National University, Busan, South Korea*

[b]*Laser Advanced System Industrialization Center, Jeonnam Technopark, Jangseong, South Korea*

[c]*Department of Mechanical Engineering, Pusan National University, Busan, South Korea*

[d]*Engineering Research Center for Net-Shape and Die Manufacturing (ERC/NSDM), Pusan National University*, Busan, South Korea

*Tel: 82-51-510-3310*

*Fax: 82-51-512-1722*

*E-Mail: bosung@pusan.ac.kr,*

Abstract

A new approach for fabricating open and closed porous structures based on laser processing is presented. Liquid polyimide (PI) was mixed with azodicarbonamide (ADC, ADA) which is a chemical blowing agent (CBA) and mixture was pre-cured in order to fabricate solid PI film. Porous PI was prepared by irradiating PI film mixed with azodicarbonamide. The PI film with azodicarbonamide was etched by laser ablation, and the CBA was decomposed by heat induced by the absorbed laser energy. The higher the laser beam irradiation, the more pores were fabricated due to the resulting increase in CBA decomposition from 27 mJ/cm$^2$ to 40 mJ/cm$^2$ per single pulse. Higher fluence at about 50 mJ/cm$^2$ resulted in fewer and larger open pores, which were formed by the coalescence of small pores. In contrast, a closed porous structure was fabricated at a fluence of less than 1 mJ/cm$^2$ because PI was barely etched. The proposed method can be sued to create open and closed micro porous structures selectively, and is not limited to thermosetting polymers, but is also effective with thermoplastic polymers.

1. Introduction

Among polymers, polyimide (PI) is well known for its excellent properties, including flexibility, low dielectric constant, good biocompatibility, outstanding mechanical properties, good thermal properties, and good chemical resistance [1]. Because of these characteristics, PI was initially applied in the space and automotive industries, and is presently used as an alternative material in conventional areas such as sensors [2], sound absorption [3], electronics and so on. Recently, PI has been investigated for application in flexible displays [4]. However, raw PI is not suitable for various products, and so researchers have studied methods to improve its properties by changing the structure and chemical formula. Among the various methods investigated, the creation of a porous structure makes it possible to reduce the weight and produce a high specific surface area that can make this material suitable for use in new applications [5].

The extrusion and injection molding method using a chemical blowing agent (CBA) is one of the conventional methods for producing a porous thermoplastic polymer [6], [7]. However, porous PI cannot be prepared using these methods because it is a thermosetting polymer. Currently several methods have been studied to fabricate porous PI structures including Ar ion beam [8], nano wire template [9], phase inversion [10], PEO-POSS nano particle [11], and salt-doped [2] methods, among others. However, these approaches are not suitable for micro scale circuits and require excessive processing time. In addition, it is very difficult to selectively fabricate an open or closed porous structure.

In this study, we prepare a PI porous structure using a laser. PI irradiated by a laser shows different phenomena at different wavelengths. It is known that the 355 nm wavelength of a UV nanosecond laser can be strongly absorbed on the PI surface; the absorption coefficient of polyimide is 20000 cm$^{-1}$ and its reciprocal absorption depth is 0.00005 cm. A 355 nm pulsed laser beam casues the effects of photothermal and photochemical ablation which generate etching of the PI [12], [13].

To the best of the author's knowledge, no method of fabricating closed and open porous PI structures on a metal, polymer, or ceramic substrate has yet been reported. In this work, a porous thermosetting PI film was fabricated using azodicarbonamide (ADC, ADA), which is a chemical blowing agent (CBA) and a 355 nm laser. The effect of laser fluence on the number of pores, the pore morphology, and the mechanism for producing closed or open porous structures was investigated in this paper.

## 2. Experimental

The liquid PI used was VTEC$^{TM}$ PI-1388 and the CBA (azodicarbonamide) was the Cellcom - AC series made by Kumyang Co., Ltd. The ADC had a gas quantity of 280-300 ml/g, and a decomposition temperature of 201-205 ℃. The laser used in the experiment was a Series 3500 UV Laser produced by DPSS Lasers Inc.: this was an Nd:YVO$_4$ 355 nm nanosecond pulse laser with an average power of 2.0 W, pulse duration of 20 ns, and TEM$_{00}$ beam mode. Figure 1 provides a schematic of the laser system.

Liquid PI mixed with the ADC was spin-coated on the PI film substrate and mixture was then pre-cured in an oven at 130 ℃ for 10 minutes. Figure 2 provides a schematic of the porous structure after six-times 355 nm pulsed laser treatment. Figure 2 (a) shows the polyimide with ADC before the laser processing. Figure 2 (b) shows the micro porous structure after the laser processing. For observation with scanning electron microscopy (SEM), the samples were coated with platinum.

## 3. Results and discussion

### 3.1 Effect of laser fluence on the number of pores

The number of pores produced in the PI film as a function of laser fluence is illustrated in Fig. 3. When the PI was irradiated with the UV laser, the higher the fluence was, the more pores the PI had per unit of surface area. One reason for this is that the at a higher fluence from 27 mJ/cm$^2$ to 40 mJ/cm$^2$ was enough to increase the PI film temperature which sufficiently activated the ADC, which changed its state to gas; and, as a thermo-chemical reaction proceeded, more more pores were finally fabricated over a short processing time. Another reason for the pore increase from 27 mJ/cm$^2$ to 40 mJ/cm$^2$ is that the 355 nm pulsed laser caused PI etching in the depth direction. Heat induced by the laser decomposed the ADC, but not many fabricated pores revealed on the surface. However, the laser beams at higher fluence help the pores produced on the surface. Higher fluence at about 50 mJ/cm$^2$ resulted in fewer and larger open pores, which were formed by the etching of the PI surface and the adding up of small pores. Although it was known that the threshold fluence between the PI and the 355 nm laser is 100 mJ/cm$^2$, it was possible to etch the the surface at a fluence lower than the ablation threshold fluence [13]. A higher fluence caused more etching of PI and revealed the closed pores existing near the surface, and added up each pores.

### 3.2 Effect of laser fluence on the pore morphology

Figure 4 provides SEM micrographs of the porous PI films for different level of laser fluence. The samples in Fig. 4(a) and Fig. 4(b) had relatively large pores. A small number of CBAs were decomposed and expanded without interference with each other at low fluence. On the other hand, Fig. 4(c) shows small pores owing to the interference between the pores that were formed by decomposition of a large number of CBAs. Compared with Fig. 4(c), Fig. 4(d) [14] shows that the volume and the number of pores increased and decreased, because the pores tended to combine with each other. And the pores gradually appeared to be enlarged by laser etching. A few closed pores were still found because the surface of the closed pores was not eliminated at low fluence as can be seen in Fig. 4(a), (b) and (c).

### 3.3 Fabrication of closed and open pore

It was found that a 355 nm laser was able to etch PI and to raise the temperature of PI above the threshold fluence. As for the etching results, gas and solid residues were produced near the etching area [15]. Figure 5 also provides an SEM micrograph of the solid residue near the pores. The wavelength of 355 nm which was generally generated by the decomposition of CBA, was almost absorbed on the thin outer surface of the closed pores; then, the thin outer surface of the closed pores decomposed by the effect of photochemical and photothermal ablation. Figure 6 provides an the SEM micrograph of a closed pore and an open pore on the PI surface at 1 $mJ/cm^2$, which value was not sufficient to eliminate the entire thin outer surface of the closed pore. That is, a small number of pores formed and solid residues barely existed near the pores. Based on these data, it was considered that closed porous structure was fabricated at a fluence of less than 1 $mJ/cm^2$.

### 4. Conclusion

A new approach for selective fabrication of open and closed porous structures on the surface of PI film, using a combination of CBA and laser irradiation has been presented. From these experimental results, some conclusions can be suggested as the follows. First, at a higher laser beam irradiation, more pores were fabricated due to the resulting increase in CBA decomposition from 27 $mJ/cm^2$ to 40 $mJ/cm^2$. Higher fluence at about 50 $mJ/cm^2$ resulted in fewer and larger open pores, which were formed by the etching of the PI surface and the adding up of small pores. Second, a closed micro porous structure was observed at a fluence level of less than 1 $mJ/cm^2$. It was found that a key factor was the effect of the laser fluence on the open and closed structures. Finally, when PI was processed by a 355 nm laser at the lower fluence, solid residue was generated. An advantage of this technology is that it can be applied to not only thermosetting polymers but also to thermoplastic polymers such as polypropylene, polyethylene, rubber on the metal, and polymer and ceramic substrates. Thus, this process can be used for selective porous structures which are impossible to fabricate by conventional methods. In the future, we will research the thermal sustainability, which is affected by the basic material and by CBA.

ACKNOWLEDGMENTS

This research was supported by Basic Science Research Program through the National Research Foundation of Korea (NRF) funded by the Ministry of Education (NRF-2015R1D1A3A01016057) and the National Research Foundation of Korea (NRF) Grant funded by the Korean Government (MSIP) (No.2015R1A5A7036513).

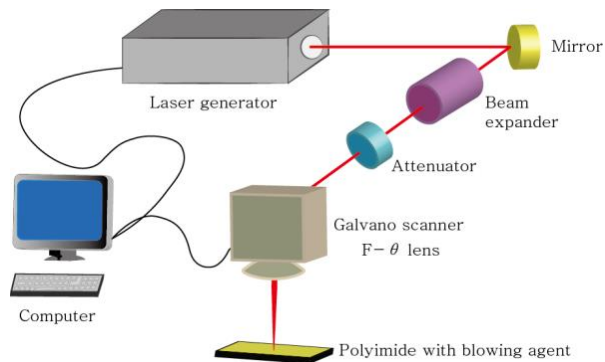

**Fig. 1.** Schematic diagram of the laser system.

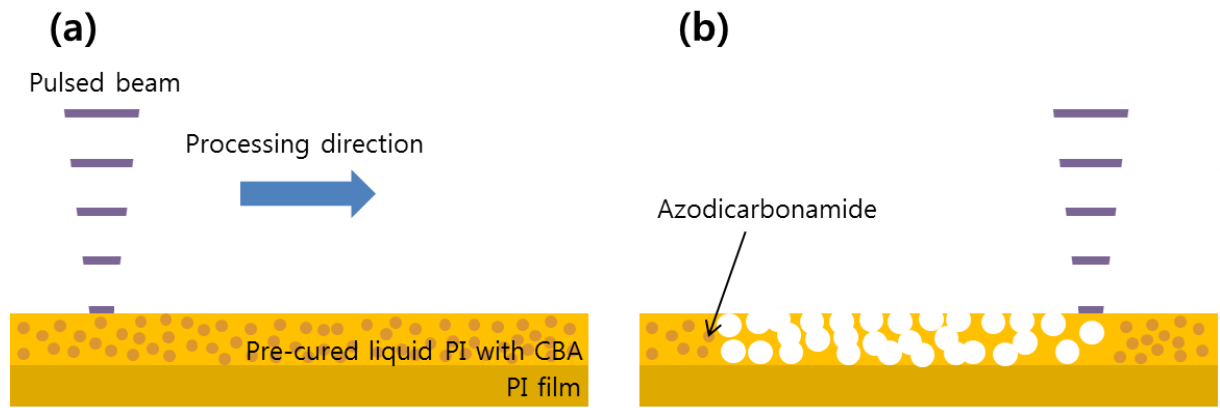

**Fig. 2.** Schematic diagram of micro porous structure fabricating process using UV-pulsed laser irradiation (a) before laser processing, (b) after laser processing.

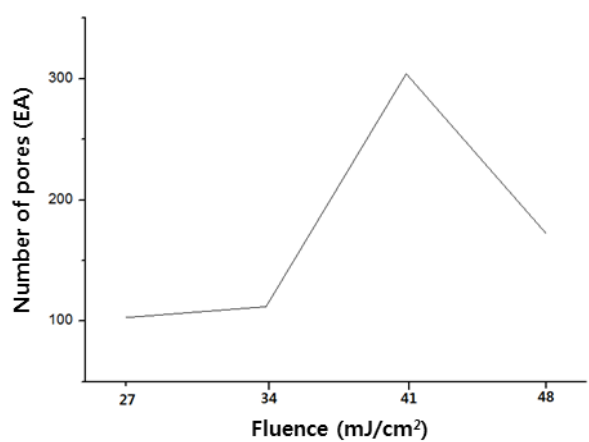

**Fig. 3.** Number of pores of PI film as a function of laser fluence.

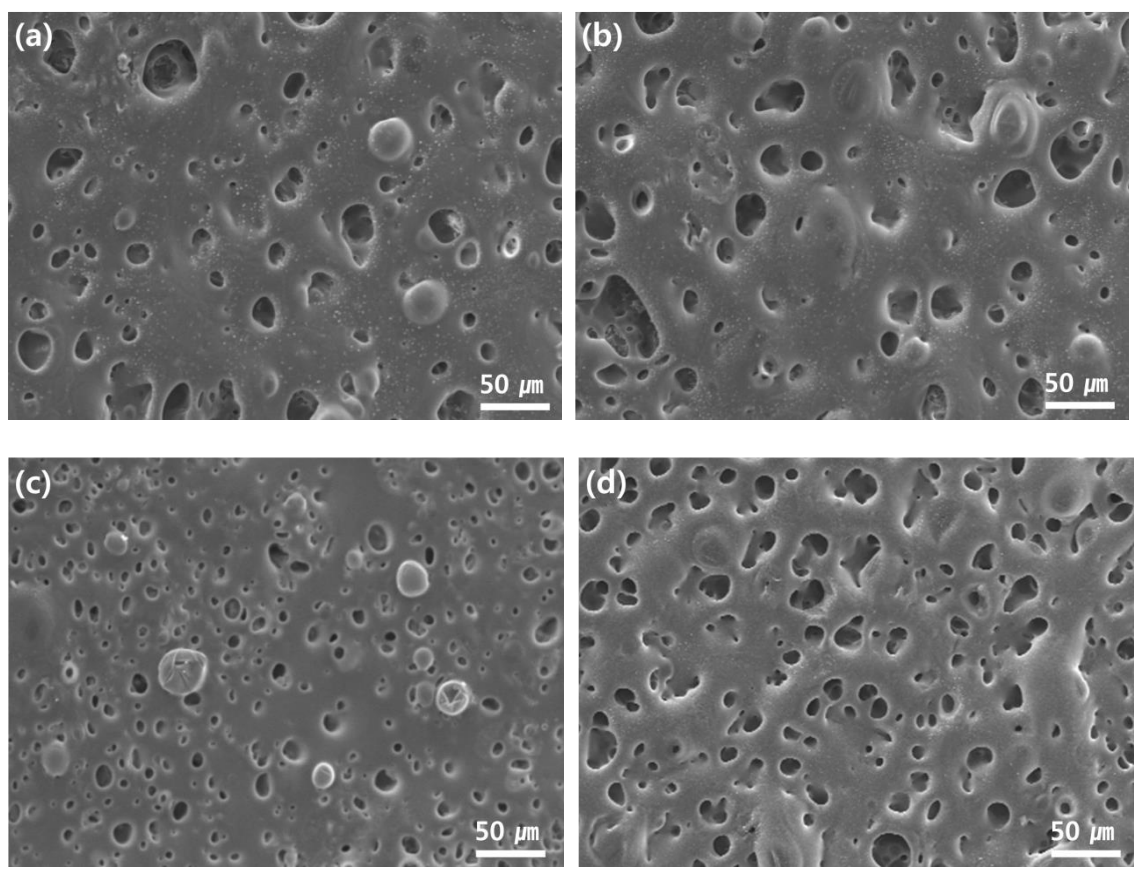

**Fig. 4.** SEM micrographs of porous PI films for different laser fluence levels: (a) 27 mJ/cm², (b) 34 mJ/cm², (c) 41 mJ/cm², (d) 48 mJ/cm².

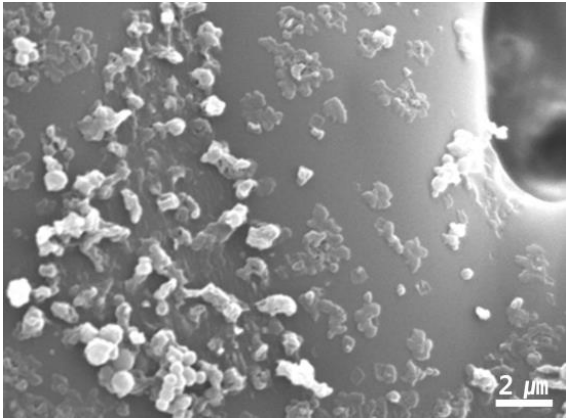

**Fig. 5.** SEM micrograph of solid residues near the pores.

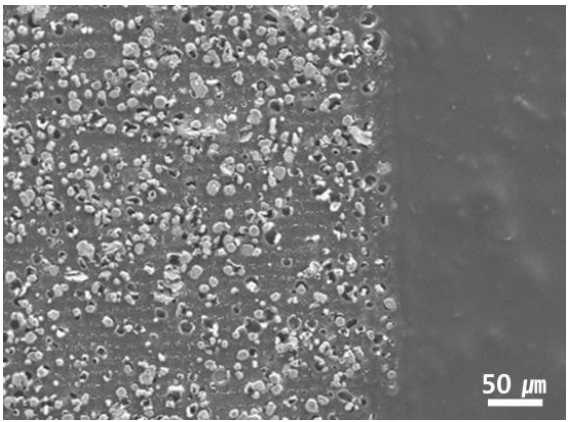

**Fig. 6.** SEM micrograph of closed pores and open pores on the PI surface at 1 mJ/cm$^2$.